\documentclass[pre,twocolumn,showpacs,preprintnumbers,amsmath,amssymb]{revtex4}

\usepackage{graphicx}
\usepackage{dcolumn}
\usepackage{bm}

\begin{document}

\newcommand{\bea}{\begin{eqnarray}}
\newcommand{\eea}{  \end{eqnarray}}
\newcommand{\bit}{\begin{itemize}}
\newcommand{\eit}{  \end{itemize}}

\newcommand{\be}{\begin{equation}}
\newcommand{\ee}{\end{equation}}
\newcommand{\ra}{\rangle}
\newcommand{\la}{\langle}
\newcommand{\U}{\widetilde{U}}


\def\bra#1{{\langle#1|}}
\def\ket#1{{|#1\rangle}}
\def\bracket#1#2{{\langle#1|#2\rangle}}
\def\inner#1#2{{\langle#1|#2\rangle}}
\def\expect#1{{\langle#1\rangle}}
\def\e{{\rm e}}
\def\proj{{\hat{\cal P}}}
\def\tr{{\rm Tr}}
\def\H{{\hat H}}
\def\Hdag{{\hat H}^\dagger}
\def\Lop{{\cal L}}
\def\Ehat{{\hat E}}
\def\Edag{{\hat E}^\dagger}
\def\Shat{\hat{S}}
\def\Sdag{{\hat S}^\dagger}
\def\Ahat{{\hat A}}
\def\Adag{{\hat A}^\dagger}
\def\U{{\hat U}}
\def\Udag{{\hat U}^\dagger}
\def\Zhat{{\hat Z}}
\def\Phat{{\hat P}}
\def\Op{{\hat O}}
\def\id{{\hat I}}
\def\x{{\hat x}}
\def\P{{\hat P}}
\def\Px{\proj_x}
\def\Pr{\proj_{R}}
\def\Pl{\proj_{L}}


\title{Weyl law for contractive maps}

\author{Mar\'\i a E. Spina}
\affiliation{Departamento de F\'\i sica, CNEA, Libertador 8250,
(C1429BNP) Buenos Aires, Argentina}
\author{Alejandro M. F. Rivas}
\affiliation{Departamento de F\'\i sica, CNEA, Libertador 8250,
(C1429BNP) Buenos Aires, Argentina}
\author{Gabriel G. Carlo}
\affiliation{Departamento de F\'\i sica, CNEA, Libertador 8250, (C1429BNP) Buenos Aires, Argentina}

\email{spina@tandar.cnea.gov.ar,rivas@tandar.cnea.gov.ar,carlo@tandar.cnea.gov.ar}

\date{\today}


\begin{abstract}

We find the Weyl law followed by the eigenvalues of contractive maps. An important property
is that it is mainly insensitive to the dimension of the corresponding invariant classical
set, the strange attractor. The usual explanation for the fractal Weyl law emergence in
scattering systems (i.e., having a projective opening) is based on classical phase space
distributions evolved up to the quantum to classical correspondence (Ehrenfest) time. In the
contractive case this reasoning fails to describe it. Instead, we conjecture that the support
for this behavior is essentially given by the strong non-orthogonality of the eigenvectors of the contractive
superoperator.

\end{abstract}

\maketitle

\section{Introduction}
\label{sec1}

The study of open quantum systems has recently become a very active field
\cite{Weiss}. The reasons are many, including the
development of quantum information and computation \cite{Nielsen,Preskill},
quantum optics and scattering systems \cite{qdot,microlaser}. Particularly in
this latter case the fractal Weyl law has been proposed. This law
predicts the way in which the long-lived resonances of these systems
grow as a function of $\hbar$. The fundamental ingredient is the
classical invariant set, which in this kind of systems is the repeller,
i.e., the set of trajectories non-escaping in the past and in the future.
In fact, this law says that the number of long-lived quasibound states
is proportional to $\hbar^{-(1+d_H)}$, where $d_H$ is the partial Hausdorff dimension
of the repeller \cite{conjecture}.

There is a vast literature that has contributed to gain confidence
on this conjecture by means of numerical tests conducted on many
systems \cite{Hamiltonians}. However, open quantum maps have been 
the main tool in these studies, as they offer great simplicity
in the calculations without losing much generality  \cite{Novaes,Nonnen,qmaps}.
 For them, the fractal Weyl law predicts that the resonances grow as
$\hbar^{-d}$, where $d$ is the partial fractal dimension of the
repeller. But if the way to open the system is nonprojective the
available literature is very scarce. Recently \cite{previous} this
situation has been analyzed for dissipative quantum operations
that can be thought as a phase space contraction leading to
dissipative dynamics \cite{contrac}. In that work a dissipative
baker map has been studied, where all classical initial conditions
asymptotically fall on a strange attractor. The quantum
counterpart has been implemented by means of a noise superoperator
written in terms of Kraus operators \cite{Kraus}. The number of
long-lived resonances has been found to behave in a rather
different way compared to the usual prediction of the fractal Weyl law. 
In fact,
this number grows as a power law in $\hbar$, but the exponent is
mainly insensitive to the dimension of the fractal invariant set.

In this work we analyze this behavior in depth.
We find the Weyl law for the spectra of contractive noise. In order to explain its emergence
and discrepancies with the usual fractal Weyl law, we
first follow the same steps than in the case of scattering systems
(i.e., having a projective opening). This is done in terms of an initial classical
distribution (that in this case shrinks following
the associated dissipation) evolved up to
the quantum to classical correspondence time $T_{\rm Ehr}$, the Ehrenfest time. We propose a
theoretical expression for this time based on dynamical considerations and confirm its
validity by means of the exploration of the classical phase space
distributions and the eigenvectors of the contractive superoperator. However, this
reasoning does not lead to a satisfactory explanation.
We conjecture that the strong non-orthogonality of the right eigenvectors
is the main reason behind this behavior.

This paper is organized as follows: in Section \ref{sec2}
we briefly describe the dissipative model that we have used and give
the expression of the Weyl law for the contractive baker map. In Section \ref{sec3} the numerical
results are analyzed and we explore possible explanations for the emergence of the here obtained Weyl law
supported by studies of the phase space distributions and the properties of eigenvectors.
Finally, we give our conclusions in Section \ref{sec4}.

\section{The Weyl law for contractive maps}
\label{sec2}

 As in our previous work \cite{previous} we have
investigated the spectral behavior of the dissipative baker map,
which is defined on the 2-torus $\mathcal T^{2}=[0,1)$ x $[0,1)$
by
\begin{equation}
\mathcal B(q,p)=\left\{
  \begin{array}{lc}
  (2q,\epsilon \: p/2) & \mbox{if } 0\leq q<1/2 \\
  (2q-1,(\epsilon \: p+1)/2) & \mbox{if } 1/2\leq q<1\\
  \end{array}\right..
\label{classicalbaker}
\end{equation}


Besides contracting the torus in the $p$ direction by a $\epsilon$
factor, this map stretches the unit square by a factor of two in the $q$
direction, squeezes it by the same factor in the $p$ direction,
and then stacks the right half onto the left one. As a result a strange
attractor sets in after a few time steps regardless of the nature of the
initial condition.

The first step to quantize it is to impose on any state $\ket{\psi}$ periodic
boundary conditions on the torus, for both the position and momentum representations.
Then, we take $\bracket{q+1}{\psi}\:=\:e^{i 2 \pi \chi_q}\bracket{q}{\psi}$, and
$\bracket{p+1}{\psi}\:=\:e^{i 2 \pi \chi_p}\bracket{p}{\psi}$,
with $\chi_q$, $\chi_p \in [0,1)$. There is a finite dimension $N=(2 \pi \hbar)^{-1}$
for the corresponding Hilbert space and a discrete set of position and momentum
eigenstates, which is given by
$\ket{q_j}\:=\:\ket{(j+\chi_q)/N}\;(j=0, 1,  \dots N-1)$, and
$\ket{p_k}\:=\:\ket{(k+\chi_p)/N}\;(k=0, 1,  \dots N-1)$,
whose eigenvalues are $q_j$, $p_k$.
A discrete Fourier transform, i.e. $$(G
_N)_{kj}  \: \equiv \: \bracket{p_k}{q_j} \:=\: \frac{1}{\sqrt{N}}
 \exp ( \frac{-i 2 \pi}{N} (j+\chi_q)(k+\chi_p)).$$ relates these sets.
We take anti-symmetric boundary
conditions, this meaning $\chi_q=\chi_p=1/2$. For an even $N$-dimensional Hilbert space,
the quantum baker map is defined in the momentum representation as
\cite{Saraceno1,Saraceno2}
\begin{equation}
\label{quantumbaker}
 B_{N}= \left(\begin{array}{cc}
  G_{N/2} & 0 \\
  0 & G_{N/2}\\
  \end{array} \right)G_{N}^{-1},
\end{equation}
with
$B_{N}$ a unitary matrix (closed quantum baker map).

We introduce dissipation by means of a non-unital quantum operation \cite{contrac}
implemented by an $ N^2 \times N^2 $ Kraus
superoperator of the form:
\begin{equation}
M=\sum_{\mu = 0}^{N-1} A^{\mu}\otimes A^{\mu \dag}.
\end{equation}
Quantum operations act on the density matrix, $\otimes$ denotes the place where this later must be inserted in order to implement the corresponding quantum operation.
Here
\begin{equation}
A^{\mu}=\sum_{i=\mu}^{N-1} \sqrt{\left( \begin{array}{c} i\\
                          i- \mu
                          \end{array}
\right) \epsilon^{i-\mu} (1-\epsilon)^{\mu}} \ket{p_{i-\mu}} \bra{p_i}
\end{equation}
are operators that induce transitions towards the momentum
state $\ket{p_{i=0}}$.
 The coupling constant $ \epsilon $ has the same value as
the dissipation parameter of the corresponding classical map.
$M$ is a trace preserving ($\sum_{\mu} A^{\dag}_{\mu}A_{\mu} = 1 $)
and non-unital ($\sum_{\mu} A_{\mu} A_{\mu}^{\dag} \neq 1 $)
superoperator, which describes a process contracting phase space
volume. The complete quantum dissipative dynamics is obtained by
composing $M$ with the unitary map (\ref{quantumbaker}),
\begin{equation}
\label{superoper}
\$= (B_N\otimes B_N^{\dag}) \circ M.
\end{equation}

\begin{figure}[h]
\begin{center}
\includegraphics*[width=1.1\linewidth,angle=0]{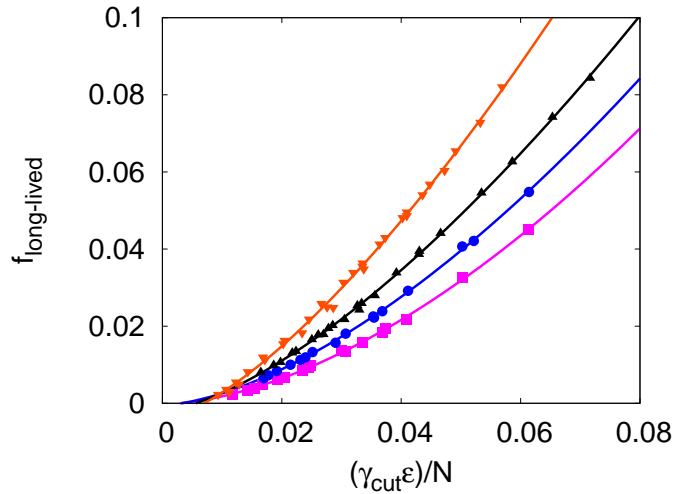}
\caption{(Color online) Weyl law for contractive noise:
 $f_{\rm  long-lived}$ as a function of $(\gamma_{\rm cut}\epsilon)/N$. Results for $\epsilon =  0.8,\ 0.7,\ 0.6$ and $ 0.4 $ are represented by means of up triangles (in red), down triangles (in black), dots (in blue) and squares (in magenta), respectively.}\label{f1}
\end{center}
\end{figure}

In this work we have computed the eigenvalue spectrum of
superoperoperator (\ref{superoper}) for different values of the
contraction parameter ($ \epsilon = 0.8,\ 0.7,\ 0.6,\ 0.4$) and of the
dimension  ($90 \leq N \leq 180$). For each case we have counted
the number of  complex eigenvalues $\lambda $ (with $|\lambda|=
\exp ({-\gamma \over 2})$) with a decay rate $ \gamma $ smaller
than a given value $\gamma_{\rm cut}$. The data are collected in Fig.
\ref{f1} which displays the fraction of resonances $
f_{\rm long-lived}$ as a function of
 $\epsilon$, $ N $ and the cut-off value
$\gamma_{\rm cut}$ (in a wide range $2 \leq \gamma_{\rm cut} \leq 14$).

By fitting these numerical results, we obtain a remarkably compact
and simple expression:

\begin{equation}
f_{\rm long-lived}(\epsilon,\gamma_{\rm cut},N) =
{N_{\gamma<\gamma_{\rm cut}} \over N^2}= C\  (\epsilon \gamma_{\rm
cut})^{2\nu} (N^2)^{-\nu}. \label{NewWeylLaw}
\end{equation}

The values of $C$ and $\nu$, for four different values of
$\epsilon$, are given in Table \ref{tab:weyl}. In the fourth
column we display the semiclassical prediction $\nu_{\rm sc}$,
which will be analyzed in Section \ref{sec3}.

\begin{center}
\begin{table}[h]
  \caption{ Values of the fitted coefficients $C$ (column two)
and $\nu$ (column three) for different values of $\epsilon$. 
The fourth column displays 
the semiclassical prediction $\nu_{\rm sc}$ described in Section \ref{sec3}. }
\label{tab:weyl}
\begin{tabular}{| r || r | c | c |}
\hline
$\epsilon=0.8$ & $C=5.3$ & $\nu=0.72$ & $\nu_{\rm sc}=0.24$\\
\hline
$\epsilon=0.7$ & $C=4.4$ & $\nu=0.76$ & $\nu_{\rm sc}=0.34$\\
\hline
$\epsilon=0.6$ & $C=4.7$ & $\nu=0.79$ & $\nu_{\rm sc}=0.42$\\
\hline
$\epsilon=0.4$ & $C=5.3$ & $\nu=0.85$ & $\nu_{\rm sc}=0.57$\\
\hline
\end{tabular}
\end{table}
\end{center}
These findings generalize the ones obtained in \cite{previous}. On
the one hand they confirm the existence of a power law dependence
of $ N_{\gamma<\gamma_{\rm cut}} \over N^2 $ on $ N $ with an exponent
which, in a meaningful range of validity, is fairly insensitive to
the value of the dissipation parameter $\epsilon $. On the other
hand, they hint (within a precision of $ 20 \% $) on a very simple
dependence of the prefactor with both $\epsilon$ and the cut-off
value $\gamma_{\rm cut}$. We will leave the analysis of this
prefactor, which is in general believed to be system-dependent,
for future work \cite{future} and concentrate in the following on
the scaling of $ N_{\gamma<\gamma_{\rm cut}} \over N^2 $ with $ N $.
We will seek for an expression of $\nu$, in order to determine to
which extent this exponent can be related to the underlying
classical dynamics. For this we will follow an approach analogous
to the one used in the formulation of the fractal Weyl law for
chaotic maps with a projective opening \cite{qmaps} and discuss its
limitations in the case of a contractive noise.

\section{Classical and quantum support for the eigenvalue statistics}
\label{sec3}

A heuristic formulation of the fractal Weyl law for chaotic maps
with a projective opening is based on the assumption that the
number of long-lived resonances (associated with the classical
repeller) scales as the volume of the evolved initial classical
distribution up to the Ehrenfest time, that is, the volume of a
finite (Ehrenfest)time repeller \cite{qmaps}. This volume can be
calculated by a combination of two exponential laws that relate
the probability to reside in the system (non escaping
trajectories) and the quantum to classical correspondence.

In the case of a contractive noise the connection between the
long-lived resonances and the structure of the classical invariant
also exists. In particular  we have verified in
\cite{previous} that the Husimi representation of the projector
corresponding to the eigenfunctions with slow escape rate
concentrates on the phase space region corresponding to the
classical strange attractor. It seems then natural to generalize
the considerations usually applied to chaotic maps with a
projective opening to the contractive case and investigate whether
this scheme succeeds in accounting for the Weyl law of
eq.(\ref{NewWeylLaw}). Our starting point will be the following
relation \cite{fwl}:

\begin{equation}
f_{\rm long-lived}(\epsilon,\gamma_{\rm cut},N) \sim  A_{\rm
clas}^2\label{propor}
\end{equation}
between the fraction of long-lived resonances and the volume of
the attractor $A_{\rm clas}$ which shrinks exponentially until the
Ehrenfest time according to:

\begin{equation}
A_{\rm clas}=\exp{-(\gamma_{\rm cl} \,\, T_{\rm Ehr})}.
\label{propor2}
\end{equation}

Notice in eq.(\ref{propor}) the  square (instead of linear)
dependence on $A_{\rm clas}$ , which is due to the use of the
superoperoperator formalism to model the contractive noise. The
classical decay rate $\gamma_{\rm cl}$ and the correspondence
(Ehrenfest) time $T_{\rm Ehr}$ are then the two main
ingredients of this approach that should be evaluated.

The classical decay rate can be easily calculated by following the
time evolution of a uniform distribution in phase space under the
action of dissipation. It is straightforward to see that after t
time steps the original distribution will occupy $2^t$ fringes in
the $q$ direction, each fringe having a width
$(\frac{\epsilon}{2})^t$. Hence the total phase space area
occupied by the distribution as a function of time is $A_{\rm
clas} \equiv  e^{-t\gamma_{\rm cl}}= \epsilon ^t$, and then  the
classical decay rate is given by $\gamma_{\rm cl}= -ln{\epsilon}$.

Determination of the Ehrenfest time is a more subtle issue.
Understood as the time at which the quantum and the classical
descriptions differ, we can start our reasoning following the
lines of what is done in the case of area preserving maps. In
fact, there are two different ways to conceive this correspondence
time. The first one is the time $T_{\rm Ehr1}$ at which a given
initial semiclassical distribution (a coherent state of width
$\sqrt{\hbar}$, for instance) spreads up to the border of the
system along the unstable direction (manifold). This time is
related to the expansive Lyapunov exponent $\lambda_1$, such that
$T_{\rm Ehr1} \propto \frac{ln{N}}{\lambdạ_{1}}$. On the other
hand, the time $T_{\rm Ehr2}$ is that corresponding to the initial
distribution shrinking along the stable direction to a size of the
order of the Planck cell ($1/N$). This time is related to the
contractive Lyapunov exponent $\lambda_2$, such that $T_{\rm Ehr2}
\propto \frac{ln{N}}{|\lambdạ_{2}|}$. Of course, in the case of
an area preserving map $\lambdạ_{1} + \lambdạ_{2} = 0$ and
both times coincide.  However, under a contractive noise,
our dissipative map gives $\lambdạ_{1}=ln{2}$ while
$\lambdạ_{2}=-ln{\frac{2}{\epsilon}}$. Hence, we propose the
shortest $T_{\rm Ehr2} \propto {ln{N}}/{ln{\frac{2}{\epsilon}}}$
as the global quantum to classical correspondence time for this
map.

\begin{figure}[h]
\begin{center}
\includegraphics*[width=0.9\linewidth,angle=0]{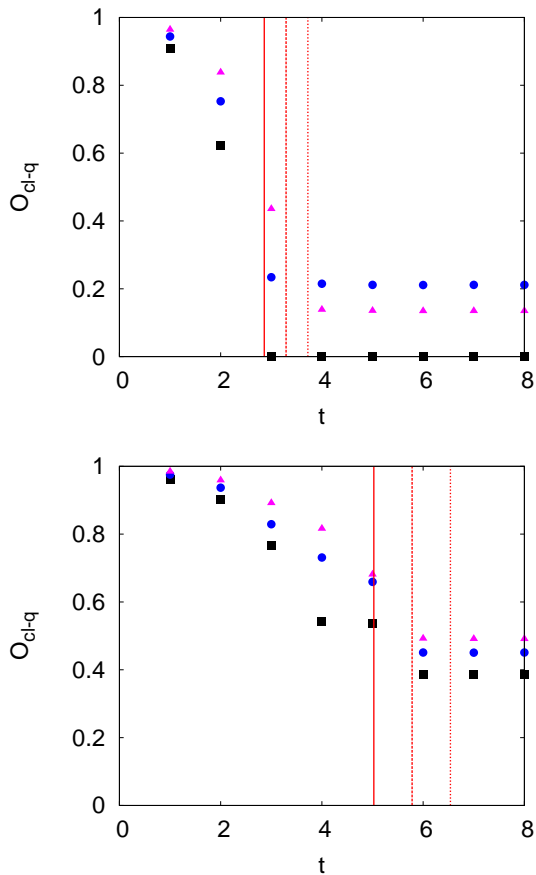}
\caption{(color online) Overlap between the phase space region
occupied by the quantum and classical attractors as a
function of time $t$ (map iterations). Upper panel corresponds to
 $\epsilon=0.4$, lower panel to $\epsilon=0.8$. Results for $N= 100,\ 200$, and $400$ are represented with squares (in black), dots (in blue) and  up triangles (in magenta), respectively. Vertical lines show
the corresponding theoretical values
of $T_{\rm Ehr2}$ for $N= 100,\ 200$, and $400$. (full, dashed and dotted lines, respectively).}
\label{f2}
\end{center}
\end{figure}

In order to verify this assumption, we have numerically estimated
the correspondence time. This can be accomplished quite easily by
evaluating the overlap $O_{\rm cl-q}$ between the finite time
classical attractor and the Husimi distribution of a uniform
initial state evolved up to the same time. If we exploit the fact
that for the baker map the interesting features of the
distribution (namely its fractality) are only in the $p$
coordinate we can notably simplify this task. In fact, we just
calculate the norm of the evolved wavefunction, restricted to the
region occupied by the classical distribution at any given time.
As a result we have obtained Fig. \ref{f2} where these overlaps
are shown as a function of the map iterations. We have found that,
besides small fluctuations and the lack of precision inherent to
the discrete time steps of the map, the results confirm our
theoretical prediction (see vertical lines as a guide).

Inserting the expressions of $\gamma_{\rm cl}$ and $T_{\rm Ehr2}$
in eq.(\ref{propor2}) gives $A_{\rm clas}=N^{-\nu_{\rm sc}}$,
where $\nu_{\rm sc}=2-d$, and
$d=1+\ln{(2)}/(\ln{(2)}-\ln{(\epsilon)})$ is the fractal dimension
of the classical attractor. The values of the semiclassical
$\nu_{\rm sc}$ are listed in the fourth column of Table
\ref{tab:weyl}, showing a dramatic discrepancy with the values
obtained by fitting our numerical results with eq.
(\ref{NewWeylLaw}). Besides an overall factor of $ \sim 2 $
between $\nu$ and $\nu_{\rm sc}$, the semiclassical exponent shows
a dependence on $\epsilon$ (via the fractal dimension of the
attractor) which is absent in the fitted $\nu$ which are
practically constant. Then, it becomes clear that the way of
reasoning that has provided with a reasonable explanation for the
emergence of the usual fractal Weyl law for systems subjected to
projective noise can no longer be applied to contractive dynamics.
We are now faced with the question of where this discrepancy comes
from.

\begin{figure}[h]
\begin{center}
\includegraphics*[width=0.9\linewidth,angle=0]{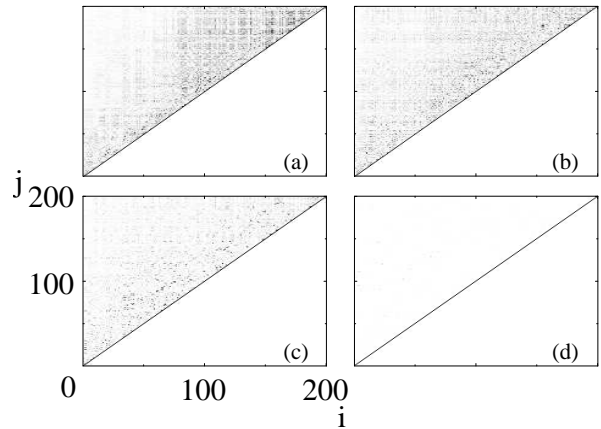}
\caption{Overlap matrices $P_{ij}$ of the first $200$ right
eigenstates with  $N=180$ (ordered by decreasing modulus of the eigenvalues) corresponding to the contractive map for $\epsilon=0.4,\
0.6$, and $0.8$, (panels (a), (b), and (c) ,respectively). 
For comparison we show the same overlap matrix but
for a projective opening that amounts to $0.2$ of the phase space.
Only the 
upper half of the matrices is shown.} 
\label{f3}
\end{center}
\end{figure}

At the basis of eq.(\ref{propor}) is the assumption that the
number of long-lived quantum states can be approximated by the
number of Planck cells which fit into the phase space volume of
the classical invariant set. This, in turn, supposes that to a
good approximation the eigenfunctions supported by this set are
non-overlapping. Even though we cannot strictly speak of
orthogonality, since the operators describing open systems are not
normal, we know that in the case of projective openings the
long-lived eigenfunctions are quasi-orthogonal (while the
short-lived ones present a high degree of degeneracy). This
explains the success of the fractal Weyl law in the projective
case. In the case of contractive dynamics we will investigate this
point by defining the overlap matrix $P_{ij}=Tr(R_i^\dag R_j)$,
where $R_i$ are the right eigenstates corresponding to the
superoperator $\$$ of Eq. (\ref{superoper}) (this is not to be
confused with the biorthogonality of the right and left
eigenfunctions of a superoperator, which states that $Tr(L_i^\dag
R_j)= \delta_{i,j}$). The overlap matrix elements corresponding to
the contractive map with $ N=180$ and $\epsilon=0.4,\ 0.6,\ 0.8$ for
the $200$ longest-lived eigenstates are displayed in panels (a),
(b), and (c) of Fig. \ref{f3}, respectively. A grayscale is used
to represent them , going from white corresponding to value $0$ to
black corresponding to the maximum values. We observe that the
off-diagonal elements are clearly non negligible. Moreover, their
value grows with the contractive power of the corresponding map
(as $\epsilon$ decreases). For comparison we show in panel (d) the
overlap matrix for a projective case, obtained by opening the
baker map along two symmetric bands in the q-direction, of width $
\delta p = 0.1$ and centered at $ p=0$ and $ p=N-1$. In this case,
as expected, the matrix is almost diagonal.

\begin{figure}[h]
\begin{center}
\includegraphics*[width=0.9\linewidth,angle=0]{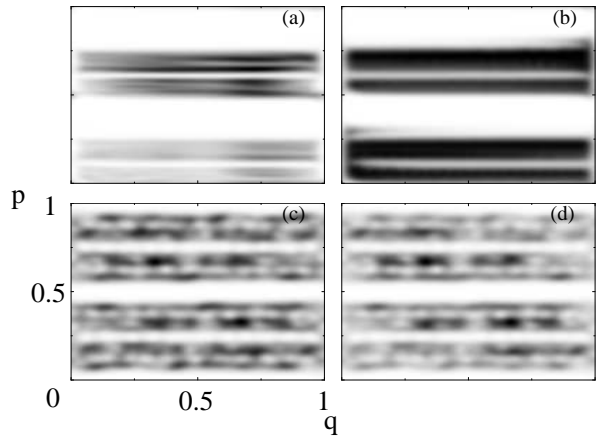}
\caption{In panel (a) we show the sum (\ref{projector})  of the Husimi representation  corresponding to the first 600 right-right eigenvectors for the contractive map at
$\epsilon=0.6$ for $N=180$. Panel (b) displays the analogous sum corresponding to the
Husimi representation of the Schur eigenvectors. 
For comparison, in the lower panels (c) and (d) we show the same distributions than in panels (a) and (b) respectively but for the projective opening case of Fig. \ref{f3} (d).}
\label{f4}
\end{center}
\end{figure}

The different degree of non-orthogonality of the long-lived
resonances in both models is also reflected in the phase space
distribution of these states. In panel a) and c) of Fig. \ref{f4} we show
the sum up to $ \gamma_{\rm cut}$ of the Husimi representation of
the longest-lived right eigenstates:
\begin{equation}
\sum_{\gamma=0}^{\gamma_{\rm cut}}{\bra{z}R_{\gamma}\ket{z}
\bra{z} R_{\gamma}^\dag\ket{z} \over \bracket{R_{\gamma}}
{R_{\gamma}^\dag}}, \label{projector}
\end{equation}
with $ \bra{z}R_{\gamma}\ket{z}=
Tr(R_{\gamma}^{\dag},\ket{z}\bra{z}) $ where $\ket{z}$ are
coherent states centered at $z=(q,p)$.
Panels b) and d) display the analogous sum (\ref{projector}) corresponding to the
Husimi representation but of the Schur eigenvectors, 
which constitute the orthogonal basis associated with the eigenvalues $ \lambda $
with $|\lambda| \geq \exp ({-\gamma_{\rm cut} \over 2})$.

In the case of the contractive map (upper line) we observe that the area of phase space occupied by the sum of the Husimi
distributions is smaller than the area corresponding to the
subspace spanned by the Schur decomposition. This is a clear sign
of the non-orthogonality of the eigenstates for this kind of
superoperators \cite{shur}.

On the contrary,  the lower panels (c) and (d) show that for the
case of a projective opening  both distributions look much the
same, indicating that the assumption of quasi-orthogonality for
the long-lived eigenfunctions is justified.

\section{Conclusions}
\label{sec4}

We have found an expression of the Weyl law for the spectra of the contractive baker map.
An analogous simple dependence on $(\gamma_{\rm cut}\epsilon)/N$ has also been obtained for a dissipative kicked top map on the sphere.
 We were not able to explain
the emergence of this law by means of the usual line of reasoning applied to the projective
case. Very simply put, the idea is counting resonances. This has been traditionally accomplished
by partitioning the phase space volume occupied by a finite time classical invariant set (the repeller).
In fact, implies a pseudo orthogonality of the long-lived eigenstates. We
could verify that this is indeed the case for the projectively opened baker map, a system that has
been paradigmatically used in the fractal Weyl law literature. But when it comes to the
dissipative baker map used in this work, we have clearly identified a high degree of non-orthogonality.
This is the main reason behind the failure of the usual reasoning for explaining the emergence of
the Weyl law.


As a result, we think that a new method for counting the long-lived resonance 
other than just partitioning the corresponding volume in phase space into Planck cells,
 is the key to understand the statistical behavior of
contractive maps. In the future, we hope to find a theoretical explanation for it, including the one
of the prefactor and the dependence on $\epsilon$ and $\gamma_{\rm cut}$ \cite{future}.

\section*{Acknowledgments}

Support from CONICET is gratefully acknowledged.

\vspace{3pc}


\end{document}